\documentclass[12pt,twoside,letterpaper,fleqn]{article}
\ifx\pdfoutput\undefined
\usepackage[pdftex,dvips,bookmarks,breaklinks]{hyperref}
\else
\usepackage{hyperref}
\fi
\hypersetup{colorlinks=false,bookmarksopen,bookmarksnumbered,citecolor=blue}

\usepackage{latexsym}

\usepackage{authblk}

\usepackage{amssymb,amsfonts,amsmath}
\usepackage{graphicx}
\usepackage{indentfirst}
\usepackage{bbm}
\usepackage{color}

\parskip=\medskipamount

\def\Real{{\mathbb R}}

\def\1{1\hspace{-4pt}1}
\def\j1{\widetilde{1\hspace{-4pt}1}}

\def\bec{\begin{center}}
\def\ec{\end{center}}
\def\a{\alpha}

\def\b{\beta}

\def\C{\Gamma}
\def\d{\delta} 

\def\e{\epsilon}

\def\r{\rho}
\def\s{\sigma}

\def\o{\omega}

\def\nn{\nonumber}

\def\ed{\end{document}}

\def\be{\begin{equation}}
\def\ee{\end{equation}}
\def\bea{\begin{eqnarray}}
\def\eea{\end{eqnarray}}
\def\ba{\begin{array}}
\def\ea{\end{array}}

\def\ft#1#2{{\textstyle{{\scriptstyle #1}
\over {\scriptstyle #2}}}}

\def\wt{\widetilde}

\begin{document}
\title{2D sigma models and differential Poisson algebras}

\author[a]{ Cesar Arias\thanks{ce.arias@uandresbello.edu}}
\author[b,c]{Nicolas Boulanger\thanks{nicolas.boulanger@umons.ac.be}}
\author[a]{Per Sundell\thanks{per.anders.sundell@gmail.com}}
\author[a,d]{ \\ Alexander Torres-Gomez\thanks{alexander.torres.gomez@gmail.com}}
\affil[a]{\textit{\small Departamento de Ciencias F\'isicas, Universidad Andres Bello-UNAB, Santiago-Chile}}
\affil[b]{\textit{\small Service de M\'ecanique et Gravitation, Universit\'e de Mons--UMONS, Belgium}}
\affil[c]{\textit{\small Laboratoire de Math\'ematiques et Physique Th\'eorique,
Universit\'e Fran\c{c}ois Rabelais,Tours, France \quad}}
\affil[d]{\textit{\small Instituto de Ciencias F\'isicas y Matem\'aticas, Universidad Austral de Chile-UACh, Valdivia-Chile}}

\renewcommand\Authands{  and }

\date{}
\maketitle
\thispagestyle{empty}

\abstract{We construct a two-dimensional topological sigma model
whose target space is endowed with a Poisson algebra for differential forms.
The model consists of an equal number of bosonic and fermionic fields of worldsheet form
degrees zero and one.
The action is built using exterior products and derivatives,
without any reference to any worldsheet metric, and is of the covariant
Hamiltonian form.
The equations of motion define a universally Cartan integrable system.
In addition to gauge symmetries, the model has one rigid nilpotent
supersymmetry corresponding to the target
space de Rham operator.
The rigid and local symmetries of the action, respectively,
are equivalent to the Poisson bracket being compatible with the de Rham
operator and obeying graded Jacobi identities.
We propose that perturbative quantization of the model yields a covariantized differential star product algebra of Kontsevich type.
We comment on the resemblance to the topological A model.
}

\newpage
\tableofcontents
\section{Introduction}

There are two ways to quantize a Poisson manifold
depending on whether the starting point is the Poisson bracket
\cite{Kontsevich} or the two-dimensional Poisson
sigma model \cite{Ikeda,Schaller}.
In the first approach, the central result is
Kontsevich's formality theorem \cite{Kontsevich}
that establishes the existence of a unique
deformation of the Poisson bracket into a
bi-differential operator, sometimes referred
to as the star product, given in an $\hbar$
expansion fixed by the conditions of
general covariance and associativity.
Inspired by string theory, Kontsevich also gave
the star product explicitly in the case of a general Poisson structure on $\Real^n$.
In the second approach, this formula was derived
using AKSZ path integral methods \cite{AKSZ} from the perturbative
expansion of the correlations functions of the Poisson
sigma model subject to suitable boundary conditions \cite{Cattaneo} (see also \cite{Cattaneo:2001ys}).
More precisely, the original works of Kontsevich and
later Cattaneo and Felder concern the deformation
of the commutative algebra of functions,
or zero-forms, on the Poisson manifold.
However, the general covariance of their 
star product formula is not manifest,
nor does it apply to differential forms
of arbitrary degrees.

A natural extension of Poisson algebras
to include higher form degrees, sometimes
referred to as differential Poisson algebras,
was defined and studied in \cite{Chu}.
Later, following the algebraic approach, the corresponding 
manifestly generally covariant form of the star
product has been studied  in \cite{Beggs,Tagliaferro,Zumino} (see also
\cite{Finnish}), though its explicit
form remains to be given beyond
$\hbar^2$-corrections\footnote{In \cite{Bursztyn},
the deformation quantization procedure has been set
up and studied at order $\hbar$ in the case of
more general vector bundles over Poisson manifolds.
}.
In this paper
we provide a generalization of the two-dimensional Poisson
sigma model in \cite{Ikeda,Schaller} as to include
fermionic worldsheet zero-forms facilitating the mapping of target space differential forms to vertex operators of worldsheet form degree zero.
We propose that its path integral quantization \textit{\`a la} Cattaneo and Felder
covariantizes Kontsevich's star product formula.

A key feature of differential Poisson algebras is
the presence of a connection one-form
$\widetilde \Gamma^\a{}_{\b}$
that is compatible with the Poisson bi-vector
$\Pi^{\a\b}$.
As we shall review in Section 2, the
covariantized Poisson bracket between
two differential forms reads
\be \{\omega, \eta\}= \Pi^{\a\b} \nabla_\a \omega \wedge \nabla_\b \eta + (-1)^{|\omega|} \Pi^{\b\gamma}\,\widetilde R^{\a}{}_{\gamma}\,i_\a\omega \wedge i_\b \eta \ ,\nn\ee
where $\nabla_\a$ has connection coefficients $\Gamma^\a_{\b\gamma}=\widetilde \Gamma^\a_{\gamma\b}$ and $\widetilde R^{\a}{}_{\b}=d\wt \Gamma^\a{}_\b+
\wt \Gamma^\a{}_\gamma \wedge \wt \Gamma^\gamma{}_\b$.
Since the connection plays a key role
in covariantizing the star product in the
algebraic approach, it is natural to seek
an extension of the original Poisson sigma model that
couples to it as well.
Another problem that one would like to address is
how to map target space $p$-forms $d\phi^{\a_1}\wedge\cdots \wedge d\phi^{\a_p}\omega_{\a_1\dots \a_p}$ to vertex operators
of form degree zero on the worldsheet.
To this end, one observes that by introducing fermionic
worldsheet zero-forms $\theta^\a$, the vertex operators
can be taken to be
$\theta^{\a_1}\cdots \theta^{\a_p}
\omega_{\a_1\dots \a_p}$ .
Thus, combining these two observations,
we are lead to adding fermionic copies $(\theta^\a,\chi_\a)$ of the original bosonic worldsheet zero- and one-forms
$(\phi^\a,\eta_\a)$. The proposed action, which we shall study in more detail
in Section \ref{Sec:3}, reads\footnote{In the Conclusions,
we shall comment on the resemblance between the action
presented here and that of the topological A model.
}
\be
S[\phi, \eta, \theta, \chi]=\int_{M_2} \left( \eta_\a \wedge d\phi^\a+\tfrac12 \Pi^{\a\b} \eta_\a \wedge \eta_\b  +\chi_\a \wedge \nabla \theta^\a +\tfrac{1}4 \Pi^{\beta \epsilon} \widetilde R_{\gamma \delta}{}^\alpha{}_\epsilon \chi_\a \wedge \chi_\b  \theta^\gamma\theta^\d \right)\ .\qquad\ \nn\ee
The role of
the additional quartic fermion coupling is to ensure a rigid supersymmetry $\delta_{\text f}$
that in particular acts as $\delta_{\text f}(\phi^\a,\theta^\a)=(\theta^\a,0)$.
Under additional conditions on the background,
the action is also invariant under gauge transformations
with one unconstrained parameter for each one-form.
We shall show that these rigid and local symmetries,
respectively, are equivalent to the bracket being compatible
with the de Rham differential and obeying graded Jacobi identities.
Thus, assuming that exist a gauge fixed action that is manifestly background
covariant, we expect that the products of the aforementioned
vertex operators contain the covariantized
Kontsevich star product for differential
forms of any degree,
whose explicit construction we leave for a future work.

The plan of the paper is as follows. In Section \ref{DPA}, we review the basic properties of
differential Poisson algebras and the conditions on the
Poisson bi-vector and curvature following from the
Jacobi identities.
In Section \ref{Sec:3}, we present the sigma model action
and show that its symmetries are equivalent to the
salient features of the differential Poisson algebra.
In Section \ref{Sec:4}, we conclude and remark
on the resemblance between our model and the
topological A model, and its potential importance
in higher spin theory.
We give our conventions and some useful identities
in Appendix \ref{conventions}.

\section{Differential Poisson algebras}\label{DPA}

\noindent In this section we recall the defining relations of Poisson differential algebras
\cite{Chu,Beggs} and the resulting form of the Poisson bracket.
The bracket consists of three compatible structures, namely a Poisson bi-vector
$\Pi^{\alpha\beta}$, a connection $\Gamma^\alpha{}_\beta$ and a one-form $S^{\alpha \beta}$.
The one-form contains the components of the bracket
that are not contained in the pre-connection \cite{Beggs},
that is, the covariant derivative along
the Hamiltonian vector field defined using $\Pi^{\alpha\beta}$.
In the symplectic case, one can set $S^{\alpha\beta}=0$ by redefining $\Gamma^\alpha{}_\beta$,
in which case the Poisson bracket is given by $\Pi^{\alpha\beta}$ and a curvature two-form
$\widetilde R^\alpha{}_\beta$ constructed from the torsion.
In what follows, we shall set $S^{\alpha \beta}=0$, leaving for future work the
analysis of whether there exists non-trivial $S$ tensors
in the non-symplectic case.

\subsection{Definition}

\noindent A differential Poisson algebra is a differential
algebra $\Omega$ endowed with a graded skew-symmetric and degree
preserving bilinear map $\{\cdot, \cdot \}$, called
Poisson bracket, that is compatible with exterior
differentiation and obeys the graded Leibniz rule,
that is
\begin{eqnarray}
\text{deg}(\{ \o_1, \o_2\}) &=& \text{deg}(\o_1) + \text{deg}(\o_2)\ , \label{P1}\\[5pt]
\{ \o_1, \o_2\} &=& (-1)^{\text{deg}(\o_1)\text{deg}(\o_2)+1}\{ \o_2, \o_1\}\ ,\label{P2}\\[5pt]
\{ \o_1, \o_2+\o_3\}&=&\{ \o_1, \o_2\} + \{ \o_1, \o_3\} \ ,  \label{P3}\\[5pt]
\{ \o_1, \o_2 \wedge \o_3 \} &=&\{ \o_1, \o_2\} \wedge \o_3 +
(-1)^{\text{deg}(\o_1)\text{deg}(\o_2)}\o_2 \wedge \{ \o_1,  \o_3 \}\ ,\label{P4}\\[5pt]
d \{\o_1,\o_2\} &=& \{d\o_1,\o_2\} + (-1)^{\text{deg}(\o_1)} \{\o_1,d\o_2\}\ ,\label{P5}
\end{eqnarray}
and that obeys the graded Jacobi identity
\bea
\{ \o_1  ,  \{ \o_2,\o_3 \}  \} &+ & (-1)^{\text{deg}(\o_1)(\text{deg}(\o_2)+\text{deg}(\o_3))}\{  \o_2 ,  \{\o_3 , \o_1\}  \} \notag \\[5pt]
&+& (-1)^{\text{deg}(\o_3)(\text{deg}(\o_1)+\text{deg}(\o_2))}\{ \o_3  ,  \{\o_1 ,\o_2 \}  \} =0\ ,\qquad
\quad \label{Jacobis}
\eea
where $\omega_i\in \Omega$ and $\text{deg}(\cdot)$ is the form degree.
We shall furthermore assume that $\Omega$
is realized as the algebra $\Omega(N)$ of differential
forms on a manifold $N$.
In what follows, we shall first use Eqs. (\ref{P1})--(\ref{P5})
to expand the Poisson bracket in terms of $\Pi$, $S$ and
the curvature $\widetilde R$, and then impose Eq. (\ref{Jacobis}).

\subsection{Poisson bi-vector and compatible connection}

\noindent Introducing local coordinates $\phi^\a$ on $N$, we define
\be
\label{Poisson}
 \Pi^{\a\b}:= \{ \phi^\a,\phi^\b\}\ ,
 \ee
which is thus an anti-symmetric tensor.
The Poisson bracket between two
zero-forms $f$ and $g$ can then
be written as
\be
\{f,g\}= \Pi^{\a \b} \partial_{\a} f \, \partial_{\b} g\ .
\ee
Next, to expand the Poisson bracket between a zero-form and a one-form
in the coordinate basis, we define
\be \Upsilon^{\a\b}:= \{\phi^\a,d\phi^\b\}=\tfrac12 d\Pi^{\a\b} +
\Sigma^{\a\b}\ ,\ee
where $\Sigma^{\a\b}$ is thus a symmetric one-form.
From the Leibniz rule, it follows that
\be \{d\phi^\a,d\phi^\b\}=d\Sigma^{\a\b}\ .\ee
Under a general coordinate transformation $\phi^\a=\phi^\a(\phi^{\prime\a'})$ with Jacobian $J^\a_{\b'}=\partial\phi^\a/\partial\phi^{\prime\b'}$ and inverse Jacobian $J^{\prime \a'}_\b$, one has the non-tensorial transformation property
\be \Upsilon^{\a\b}=J^\a_{\a'} J^\b_{\b'} \Upsilon^{\prime \a'\b'}+J^\a_{\a'}
\Pi^{\prime \a'\b'}d  J^\b_{\b'}\ .\label{sigma}\ee
Introducing a connection one-form
\be \widetilde \Gamma^\a{}_\b=d\phi^\gamma \widetilde\Gamma^{\a}_{\gamma\b}\ ,\ee
one has the transformation law
\be \Pi^{\a\gamma} \widetilde\Gamma^\b{}_\gamma=J^\a_{\a'} J^\b_{\b'} \Pi^{\prime \a'\gamma'} \widetilde\Gamma^{\prime \b'}{}_{\gamma'}-\Pi^{\a\gamma}J^{\prime \b'}_\gamma d J^\beta_{\b'}\ .\ee
Using the tensorial transformation property of
$\Pi$ to rewrite (\ref{sigma}) as
\be \Upsilon^{\a\b}= J^\a_{\a'} J^\b_{\b'} \Upsilon^{\prime \a'\b'}+\Pi^{\a\gamma} J^{\prime\b'}_{\gamma}d J^\beta_{\b'}\ ,\ee
we can thus write
\be \Upsilon^{\a\b}=U^{\a\b}-\Pi^{\a\gamma}\widetilde\Gamma^{\b}{}_\gamma\ ,\ee
where $U^{\a\b}$ is a tensorial one-form.
It follows that
\begin{eqnarray}
\Upsilon^{\a\b}_\gamma&=&\tfrac12 \partial_\gamma \Pi^{\a\b}+\Sigma^{\a\b}_\gamma , \notag \\
 &=&\tfrac12 \widetilde\nabla_\gamma \Pi^{\a\b}+\Sigma^{\a\b}_\gamma-\Pi^{\d[\b}\widetilde\Gamma^{\a]}_{\gamma\d},  \notag \\
&=&\tfrac12 \widetilde \nabla_\gamma \Pi^{\a\b}+\Sigma^{\a\b}_\gamma-\Pi^{\d(\a}\widetilde\Gamma^{\b)}_{\gamma\d}-\Pi^{\a\d}
\widetilde\Gamma^\b_{\gamma\d}\ .
\end{eqnarray}
Thus $U^{\a\b}=\tfrac12 \widetilde \nabla \Pi^{\a\b}+S^{\a\b}$, where
\be S^{\a\b}_\gamma:=\Sigma^{\a\b}_\gamma-\Pi^{\d(\a}\widetilde\Gamma^{\b)}_{\gamma\d}
\label{S}\ee
are the components of a tensorial one-form.
In summary, we can write
\be \label{0-1-PB}
\{\phi^\a,d\phi^\b
\}= \tfrac12 \widetilde\nabla \Pi^{\a\b} + S^{\a\b}- \Pi^{\a\gamma}\widetilde\Gamma^\b{}_\gamma\ ,\ee
where the first two terms are tensorial and the last term, which is non-tensorial, is sometimes referred to as the pre-connection \cite{Beggs}.

It is convenient to choose the connection to belong to
the equivalence class obeying
\be \widetilde\nabla_\a \Pi^{\b\gamma}=0\ ,\label{comp}\ee
which we shall assume henceforth.
It follows that
\be \{d\phi^\a,d\phi^\b\}=-\widetilde R^{\a\b}+\Pi^{\gamma\d}\widetilde\Gamma^\a_\gamma \wedge  \widetilde \C^\b_\d+ dS^{\a\b}\ ,\label{PB11}\ee
where the two-form
\be \widetilde R^{\a\b}:=\Pi^{\b\gamma}\widetilde R^\a{}_\gamma=\widetilde R^{\b\a}\ ,
\label{tildeR}\ee
as a consequence of (\ref{comp}).
%

\subsection{Manifestly covariant Poisson bracket}

\noindent Let $\omega$ and $\eta$ be differential forms of any
degree.
From the basic Poisson brackets \eqref{Poisson},
\eqref{0-1-PB} and \eqref{PB11} in the coordinate basis,
and invoking (\ref{P2}) and (\ref{P4}), it then follows
that
\bea \{\omega, \eta\}&=& \Pi^{\a\b} \nabla_\a \omega \wedge \nabla_\b \eta +  S^{\a\b} \left( (-1)^{|\omega|} \nabla_\a \omega \wedge i_\b \eta - i_\a \omega \wedge \nabla_\b \eta \right)\nn \\[5pt]&&+ (-1)^{|\omega|} \left(\widetilde R^{\a\b} -\widetilde\nabla S^{\a\b}\right) \wedge i_\a \omega \wedge i_\b \eta \ ,\label{PB}\eea
where $i$ denotes inner differentiation and $\nabla$ uses the connection coefficients
\be \Gamma^\a_{\b\gamma}:=\widetilde \Gamma^\a_{\gamma\b}\ .\label{Gamma}\ee
By construction, the above manifestly covariant
form of the Poisson bracket obeys \eqref{P5}, that is, it
is compatible with the de Rham operator.

The equivalence class of compatible connections
is generated by shifts $\delta \widetilde \Gamma^\a_{\b\gamma}$
obeying $\delta(\widetilde\nabla_\a \Pi^{\b\gamma})= 0$ and
$\delta \Pi^{\a\b}=0$,
that is
$\delta \widetilde \Gamma^{[\b}_{\a\d} \Pi^{\gamma]\d}=0$.
Under such shifts, the Poisson bracket
(\ref{0-1-PB}) is left invariant provided
\be \delta S^{\a\b}=\Pi^{\a\gamma}\delta \widetilde\Gamma^\b_\gamma\ ,\label{ShiftS}\ee
which is indeed symmetric in $\a$ and $\b$.
The invariance of the full Poisson bracket (\ref{PB})
can then be verified using
$\delta \nabla_\a \omega= -\delta \widetilde\Gamma^\b_\a i_\b \omega$ and $\delta \widetilde R^{\a\b} = \Pi^{\a\gamma}\widetilde \nabla \delta \widetilde \Gamma^\b_\gamma$.
In the symplectic case, the shift symmetry (\ref{ShiftS}) can be used to set $S=0$.
In what follows, we shall specialize to the case $S=0$,
which has been studied in detail in \cite{Chu, Beggs, Tagliaferro, Zumino},
leaving the analysis of the general case for future work.

\subsection{Jacobi identities}

\noindent In order to analyze the Jacobi identities (\ref{Jacobis}) (in the case
that $S=0$), one can use \eqref{P4} to show that if they hold for functions and
one-forms then they hold for forms of any degree.
In the case of three functions $f_1,f_2, f_3$, one finds
\be 0=\{f_{[1},\{f_2,f_{3]}\}\}= 3\nabla_\a f_{[1} \nabla_\b f_2 \nabla_\gamma f_{3]} \Pi^{\a\d}T^\b_{\d\e}\Pi^{\e\gamma}\ ,\ee
from which it follows that
\be J_0^{\a\b\gamma}:=\Pi^{\d[\a}T^\b_{\d\e}\Pi^{\gamma]\e}= 0\ .\ee
In view of $\widetilde\nabla_\a \Pi^{\gamma\d} = 0$,
this condition is equivalent to that $\Pi$ is a Poisson bi-vector,
\emph{i.e.}
\be
\Pi^{\d[\a} \partial_{\d}\Pi^{\b\gamma]} = 0\ .\label{J000}
\ee
In the case of two function $f_1, f_2$ and a one-form $\omega$,
the Jacobi identities read
\be 0= 2\{f_{[1},\{f_{2]},\omega\}\}+ \{\omega,\{f_1,f_2\}\}= \nabla_\a f_{[1} \nabla_\b f_{2]} \omega_\gamma \Pi^{\a\d}\Pi^{\b\e} R_{\d\e}{}^\gamma{}_\lambda d\phi^\lambda\
\ ,\ee
from which we obtain
\be J_1^{\a\b,\gamma}{}_\lambda:=\Pi^{\a\d}\Pi^{\b\e} R_{\d\e}{}^\gamma{}_\lambda= 0\label{J001}\ .\ee
Finally, for a single function $f$ and two one-forms $\omega_1, \omega_2$, we have
\be 0=\{f,\{\omega_{(1},\omega_{2)}\}\}+ 2\{\omega_{(1},\{\omega_{2)},f\}\}= -\nabla_\a f i_\b \o_{(1} i_\gamma \o_{2)} \Pi^{\a\d} \nabla_\d \widetilde R^{\b\gamma}\ ,\ee
which implies that
\be J_2^{\a,\b\gamma}{}_{\d\e}:=\Pi^{\a\lambda}\nabla_\lambda \widetilde R_{\d\e}{}^{\b\gamma}= 0\ .\label{J011}\ee
Finally, for three one-forms one finds that
\be J_3^{\a\b\gamma}{}_{\r\s\lambda}:=\widetilde R_{\epsilon [ \rho}{}^{(\alpha \beta} \widetilde R_{\sigma \lambda]}{}^{\gamma) \epsilon} = 0\ .\label{J111}\ee
As observed in \cite{Beggs,Zumino}, the compatibility between the
Poisson bracket and the de Rham differential implies that the
independent conditions are given by the following irreducible
representations\footnote{From \eqref{indep} the remaining conditions
follow by covariant differentiation, \emph{viz.}$$
J_1^{[\a\b,\gamma]}{}_\lambda\sim \nabla_\lambda J_0^{\a\b\gamma}\ ,\quad
J_2^{[\a,\b]\gamma}{}_{\d\e}\sim
\nabla_{[\d} J_1^{\a\b,\gamma}{}_{\e]}\ ,\quad J_3^{\a\b\gamma}{}_{\d\e\lambda}
\sim \nabla_{[\d} J_2^{(\a,\b\gamma)}{}_{\e\lambda]}\ .$$}
\be J_0^{\a\b\gamma}=0\ ,\qquad J_1^{\a(\b,\gamma)}{}_\lambda=0
\ ,\qquad J_2^{(\a,\b\gamma)}{}_{\d\e}=0\ .\label{indep}\ee
As for examples of non-trivial solutions, see \cite{Chu,Beggs}.

\section{Poisson sigma model}\label{Sec:3}

\noindent
In this section, we use the Poisson bi-vector
and its compatible connection to construct
the couplings in a two-dimensional topological
sigma model action that exhibits an extra nilpotent
rigid supersymmetry
$\delta_{\text f}$ corresponding to the de Rham differential on $N$.
As we shall see, the rigid symmetry fixes the coefficient of the
quartic fermion coupling while the gauge symmetries require
the background fields to obey the conditions (\ref{J000}),
(\ref{J001}), (\ref{J011}) and (\ref{J111}), which we recall
are equivalent to that the underlying differential
Poisson algebra obeys the Jacobi identities.

\subsection{The action}\label{Sec:3.1}

\noindent
Our action, which is formulated on a two-dimensional
manifold $M_2$, is given by
\be\label{action}
S=\int_{M_2} \left( \eta_\a \wedge d\phi^\a+\tfrac12 \Pi^{\a\b} \eta_\a \wedge \eta_\b  +\chi_\a \wedge \nabla \theta^\a +\tfrac{1}4
\widetilde R_{\gamma \delta}{}^{\alpha\beta} \chi_\a \wedge \chi_\b  \,\theta^\gamma\,\theta^\d \right)\ ,\qquad\ee
where $\widetilde R_{\gamma\d}{}^{\a\b}=\Pi^{\b\e} \widetilde R_{\gamma\d}{}^{\a}{}_\e$
are the components of the two-form (\ref{tildeR}) obtained from the
Poisson bi-vector
and its compatible connection, and the covariant exterior derivative
\be \nabla \theta^\a := d\theta^\a+d\phi^\b\Gamma^\a_{\b\gamma} \theta^\gamma\ ,\ee
where the connection coefficients are defined in \eqref{Gamma}.
The fields are assigned form degrees ${\rm deg}_2$ on $M_2$ and
an additional Grassmann parity $\e_{\rm f}(\cdot)$ as follows:
\be {\rm deg}_2(\phi^\a;\eta_\a,\theta^\a,\chi_\a)= (0;1,0,1)\ ,\qquad
\e_{\rm f}(\phi^\a;\eta_\a,\theta^\a,\chi_\a)=(0;0,1,1)\ .\ee
In what follows, we shall assume $M_2$ to be compact
and that the pull-backs of $(\eta_\a,\chi_\a)$ to the boundary
of $M_2$ vanish.

Geometrically, the action describes a sigma model
with source $M_2$ and target space given by the
$\mathbb N$-graded bundle 
\be \widehat N=T^\ast[1,0]N\oplus T[0,1]N\oplus T^\ast[1,1] N\ ,
\label{widehatM}\ee
coordinatized by $(\phi^\a;\eta_\a,\theta^\a,\chi_\a)$,
where $T[p,\e]N$ is the degree shift of the tangent bundle $T[0,\e]N$
over $N$ by $p$ units \emph{idem} $T^\ast[p,\e]$.
The Grassmann parity of $T[p,\e]N$ is $\e_{\rm f}(T[p,\e]N)=\e$.
The sigma model map $\varphi:M_2\rightarrow \widehat N$ has
vanishing intrinsic degree and Grassmann parity,
and the degree on $M_2$ is the form degree on $M_2$.
Thus, if $\omega_{n,p,\e}$ denotes
an $n$-form on $\widehat N$ of degree $p$ and
Grassmann parity $\e$, then its pull-back $\omega_{p,\e}:=\varphi^\ast \omega_{n,p,\e}$
is a $p$-form on $M_2$ of Grassmann parity $\e$.
The de Rham differential on $\widehat N$ has
form degree one and degree one.
We use the following Koszul sign convention, which is consistent with
Leibniz' rule:
\be \omega_{n_1,p_1,\e_1}\wedge \omega_{n_2,p_2,\e_2}=(-1)^{p_1p_2+\e_1\e_2}
\omega_{n_2,p_2,\e_2}\wedge \omega_{n_1,p_1,\e_1}\ .\ee
The resulting sign convention for wedge
products on $M_2$ reads
\be \omega_{p_1,\e_1}\wedge \omega_{p_2,\e_2}=(-1)^{p_1p_2+\e_1\e_2}
\omega_{p_2,\e_2}\wedge \omega_{p_1,\e_1}\ .\ee
The manifest target space covariance of the action amounts to the fact that
target space diffeomorphisms
\be \delta_\xi \phi^\a = \xi^\a\ ,\quad \delta_\xi \eta_\a = -\partial_\a \xi^\b \eta_\b\ ,\quad
\delta_\xi \theta^\a = \partial_\b \xi^\a \theta^\b\ ,\quad \delta_\xi \chi_\a = -\partial_\a \xi^\b \chi_\b\ ,\ee
which act on the worldsheet fields, induce Lie derivatives acting on
the background fields, \emph{i.e.}
\be \delta_\xi S[\phi,\eta,\theta,\chi;\Pi,\Gamma]={\cal L}_\xi S[\phi,\eta,\theta,\chi;\Pi,\Gamma]\ ,\label{diff}\ee
where
\bea {\cal L}_\xi \Pi^{\a\b} &=& \xi^\gamma\partial_\gamma  \Pi^{\a\b}+2 \partial_\gamma \xi^{[\a} \Pi^{\b]\gamma}\ ,\\[5pt]{\cal L}_\xi \Gamma^\a_{\b\gamma} &=& \partial_\b\partial_\gamma\xi^\a+ \xi^\d\partial_\d  \Gamma^\a_{\b\gamma}-  \partial_\d \xi^{\a} \Gamma^\d_{\b\gamma}+ \partial_\b \xi^\d \Gamma^\a_{\d\gamma}+\partial_\gamma \xi^\d \Gamma^\a_{\b\d}\ .\eea
%

\subsection{Nilpotent rigid fermionic symmetry}

\noindent  The de Rham differential $d=d\phi^\a \partial_\a$ on $N$ lifts to a holonomic vector field $\theta^\a\partial_\a$
on $T[0,1]N$, which in its turn can be extended to a
nilpotent rigid supersymmetry $\delta_{\rm f}$ of the action as follows:
\bea\label{susy}
\delta_{\text f} \phi^\alpha&=&\theta^\alpha \ , \notag \\
\delta_{\text f} \theta^\alpha&=&0 \ , \notag \\
\delta_{\text f} \eta_\alpha&=&\Gamma^\beta_{\alpha\gamma} \eta_\beta \theta^\gamma+\tfrac1 2 \widetilde R_{\beta\gamma}{}^\delta{}_\alpha\, \chi_\delta \, \theta^\beta \theta^\gamma \ , \notag \\
\delta_{\text f} \chi_\alpha&=&-\eta_\alpha-\Gamma^\beta_{\alpha\gamma} \chi_\beta  \theta^\gamma \ ,
\eea
as can be seen using $\widetilde \nabla_\alpha \Pi^{\beta\gamma}=0$  and the Bianchi identity $\widetilde \nabla_{[\alpha} \widetilde R_{\beta\gamma]}{}^\delta{}_\epsilon-\widetilde T^\lambda_{[\alpha\beta} \widetilde R_{\gamma]\lambda}{}^\delta{}_\epsilon=0$.
We note that ${\rm deg}_2(\delta_{\text f})=0$, $\e_{\rm f}(\delta_{\text f})=1$ and that
$\delta_{\text f}^2 \eta_\alpha=0$ requires the aforementioned Bianchi identity.
Moreover, just as the relative coefficient between the two terms in
the Poisson bracket (\ref{PB}) is fixed by compatibility with the
$d$ operator, the rigid supersymmetry requirement
fixes the relative strength between the kinetic terms
and the quartic fermion term in the action (\ref{action}).
In fact, the $\delta_{\text f}$-invariance of the action can
be made manifest by observing that the Lagrangian is
$\delta_{\text f}$-exact, \emph{viz.}
\begin{equation}
S\equiv \int_{M_2}  L \ ,\qquad L= \delta_{\text f} V\ ,\label{V}
\end{equation}
where
\begin{equation}
V=\ -\chi_\alpha \wedge \left( d\phi^\alpha+\tfrac 1 2 \Pi^{\alpha \beta} \eta_\beta  \right)  \ ,
\end{equation}
as can easily be seen using $\widetilde \nabla_\alpha \Pi^{\beta\gamma}=0$ and we note
that there is no need to discard any total derivative in \eqref{V}.
The commutator between the rigid supersymmetry and the target space
diffeomorphisms takes the form
\bea [\delta_\xi, \delta_{\text f}] \phi^{\a} &=& 0\ ,\qquad  [\delta_\xi, \delta_{\text f}]
\eta_\a= {\cal L}_\xi \Gamma^{\gamma}_{\a\b}\, \eta_\gamma\theta^\b +
{\cal L}_\xi \widetilde R_{\b\gamma}{}^\d{}_\a \chi_\d \,\theta^\b\theta^\gamma\ ,\\[5pt]  [\delta_\xi, \delta_{\text f}] \theta^{\a} &=& 0\ ,\qquad  [\delta_\xi, \delta_{\text f}]
\chi_\a= -{\cal L}_\xi \Gamma^{\gamma}_{\a\b} \chi_\gamma \theta^\b \ .\eea
Thus the rigid supersymmetry commutes with background Killing symmetries,
whose Lie derivatives by definition annihilate $\Pi^{\a\b}$ and $\Gamma^\a_{\b\gamma}$ and hence the action as can be seen from \eqref{diff}.

\subsection{Equations of motion}

\noindent Applying the variational principle to the action
(\ref{action}) yields the following equations of motion:
\bea
{\cal R}^{\phi^\a} &:=& d\phi^\a+ \Pi^{\a\b}\eta_\b=0\ ,\label{Rphi}\\[5pt]
{\cal R}^{\theta^\a} &:=& \nabla \theta^\a + \tfrac12 \widetilde R_{\gamma\d}{}^{\a\b}\chi_\b\theta^\gamma \theta^\d =0\ ,\label{Rtheta}\\[5pt]
{\cal R}^{\chi_\a} &:=& \nabla \chi_\a - \tfrac12  \widetilde R_{\a\d}{}^{\b\gamma}\chi_\b  \wedge \chi_\gamma \theta^\d = 0\ ,\label{Rchi}\\[5pt]
{\cal R}^{\eta_\a} &:=& \nabla \eta_\a +  R_{\a \gamma}{}^{\b}{}_{\d}\chi_\b \wedge d\phi^\gamma \theta^\d +\tfrac 1 4  \nabla_\a \widetilde R_{\d\e}{}^{\b\gamma}\chi_\b \wedge \chi_\gamma  \theta^\d\theta^\e = 0\ ,\label{Reta}
\eea
where
\be
\nabla \chi_\a :=  d\chi_\a-d\phi^\b\Gamma^\gamma_{\b\a} \wedge \chi_\gamma\ ,
\ee
\emph{idem} $\nabla \eta_\a$.
We note that Eqs. (\ref{Rphi})--(\ref{Rchi}) are given by the
functional derivatives of $S$ with respect to $(\eta_\a,\chi_\a,\theta^\a)$,
respectively, while Eq. (\ref{Reta}) has been obtained from
\bea \frac{\delta S}{\delta \phi^\a}&=& d\eta_\alpha+\tfrac1 2 \partial_\alpha \Pi^{\beta\gamma} \eta_\beta \wedge \eta_\gamma+ (\Gamma^\gamma_{\alpha \beta}d\chi_\gamma -\chi_\gamma \wedge d\Gamma^\gamma_{\alpha \beta}+\partial_\alpha \Gamma^\gamma_{\delta \beta} \chi_\gamma \wedge d\phi^\delta ) \theta^\beta \notag \\[5pt]
&-&\Gamma^\gamma_{\alpha \beta}\chi_\gamma  \wedge d\theta^\beta+\tfrac 1 4  \partial_\alpha \widetilde R_{\beta \gamma}{}^{\delta\epsilon}\chi_\delta \wedge \chi_\epsilon \theta^\beta \theta^\gamma
\ ,\eea
by rewriting $\partial_\alpha\Pi^{\beta\gamma} \eta_\beta \eta_\gamma$
using ${\cal R}^{\phi^\a}=0$ and $\widetilde \nabla_\alpha\Pi^{\b\gamma}=0$,
and the quantities $d\chi_\gamma \Gamma^\gamma_{\alpha \beta} \theta^\beta$
and $\chi_\gamma \Gamma^\gamma_{\alpha\beta} d\theta^\beta$
using ${\cal R}^{\chi_\a}= 0$ and ${\cal R}^{\theta^\a}=0$, respectively.

\subsection{Universal Cartan integrability}

\noindent Let us demonstrate that the universal
Cartan integrability of the equations of motion,
which is required for the validity of Cartan
gauge symmetries and on-shell integration,
is equivalent to that the target space background
obeys the conditions (\ref{J000}),
(\ref{J001}), (\ref{J011}) and (\ref{J111}),
\emph{i.e.} that they can be used to
define a differential Poisson algebra
obeying the Jacobi identity (\ref{Jacobis}).
To this end, we derive the generalized Bianchi identities
\be \nabla {\cal R}^i + M^i_j \wedge {\cal R}^j+{\cal A}^i = 0\ ,\ee
where ${\cal R}^i:=({\cal R}^{\phi^\a},{\cal R}^{\theta^\a},{\cal R}^{\chi_\a},
{\cal R}^{\eta_\a})$ and $M^i_j$ is a field dependent
matrix, after which we require compatibility
in the universal sense, that is, that the classical
anomalies ${\cal A}^i$ vanishes on base manifolds
of arbitrary dimensions.

As for $\nabla\mathcal R^{\phi^\a}$, and using $\nabla d\phi^\alpha=T^\alpha$,
the resulting compatibility condition reads
\bea
{\cal A}^{\phi^\a}&=&-\left(  \ft12 \Pi^{\rho \delta} T^\alpha_{\rho \epsilon}  \Pi^{ \sigma\epsilon}+ \nabla_\rho \Pi^{\alpha \sigma} \Pi^{\rho \delta} \right) \eta_\delta \wedge \eta_\sigma+ \Pi^{\alpha \rho} \Pi^{\gamma \sigma} R_{\rho \gamma}{}^\beta{}_\delta \, \chi_\beta \wedge \eta_\sigma \theta^\delta\nn\\[5pt]
&&-\tfrac 1 4 \Pi^{\alpha \rho} \nabla_\rho \widetilde R_{\delta \epsilon}{}^{\beta \gamma}\, \chi_\beta \wedge \chi_\gamma \theta^\delta \theta^\epsilon\ ,
\eea
that must thus hold without imposing any algebraic constraint on $(\phi^\a,\eta_\a;\theta^\a,\chi_\a)$.
Thus, using also the identity $\nabla_\rho \Pi^{\alpha \sigma}=-2 T^{[\alpha }_{\rho \epsilon}\Pi^{\sigma]\epsilon }$, which allows us to rewrite the first term as
$\ft32 \, \Pi^{\rho[\delta} T^{\alpha}_{\rho \epsilon} \Pi^{\sigma] \epsilon}\eta_\delta \wedge \eta_\sigma$, the vanishing of $\mathcal A^{\phi^\a}$ requires
\be
\Pi^{\d [\a}T^\b_{\d\e} \Pi^{\gamma]\e} =  0   \ , \qquad \Pi^{\alpha \rho} \Pi^{\sigma \beta} R_{\rho \sigma}{}^{\gamma}{}_\delta = 0 \ , \qquad \Pi^{\alpha \lambda} \nabla_\lambda \widetilde R_{\beta \gamma}{}^{\rho \sigma} = 0 \ ,\label{conds}
\ee
which we identify as the complete set of conditions required
for the Jacobi identity (\ref{Jacobis}).
In particular, the second condition in (\ref{conds}) is equivalent to that
\be \nabla^2=0\qquad\mbox{on-shell}\ .\label{nablasquared}\ee
Next, taking into account (\ref{conds}), the vanishing of $\mathcal A^{\theta^\alpha}=0$ requires the additional condition
\be
\widetilde R_{\epsilon \rho}{}^{(\alpha \beta} \widetilde R_{\sigma \lambda}{}^{\gamma) \epsilon}\, \chi_\beta \wedge \chi_\gamma \theta^\rho \theta^\sigma \theta^\lambda = 0\ ,
\ee
or, equivalently,
\be
\widetilde R_{\epsilon [ \rho}{}^{(\alpha \beta} \widetilde R_{\sigma \lambda]}{}^{\gamma) \epsilon} = 0\ . \label{conds2}
\ee
which is a consequence of the previous conditions,
as discussed below Eq. (\ref{J011}).
Turning to the integrability of $\mathcal R^{\chi_\alpha}=0$,
it follows from (\ref{conds}) and (\ref{conds2}) that
$\mathcal A^{\chi_\alpha}$ vanishes universally,
noting that the $\chi^{\wedge 3}\theta^2$-terms in
$\nabla\mathcal R^{\chi_\alpha}$ are proportional to
$\widetilde R_{\epsilon[ \alpha}{}^{(\rho \sigma} \widetilde R_{\beta \gamma]}{}^{\lambda)\epsilon}\, \chi_\rho \wedge  \chi_\sigma \wedge \chi_\lambda \theta^\alpha \theta^\gamma$.
Finally, using (\ref{nablasquared}) one has
\begin{eqnarray}
\mathcal A^{\eta_\alpha}=&-&\tfrac 1 4 \Pi^{\lambda\beta} ( \nabla_\beta \nabla_\alpha \widetilde R_{\gamma \delta}{}^{\rho \sigma}+2 R_{\alpha \beta}{}^\rho{}_\epsilon  \widetilde R_{\gamma \delta}{}^{\sigma\epsilon }+ 2 R_{\alpha \beta}{}^\epsilon{}_\gamma \widetilde R_{\delta\epsilon }{}^{\rho \sigma} ) \chi_\rho \wedge \chi_\sigma \wedge \eta_\lambda \theta^\gamma \theta^\delta \notag \\
&+&\tfrac 1 4 \nabla_\alpha(\widetilde R_{\beta \delta}{}^{\rho \sigma} \widetilde R_{\epsilon \gamma}{}^{\lambda \beta }) \, \chi_\rho \wedge \chi_\sigma \wedge \chi_\lambda \theta^\delta \theta^\epsilon \theta^\gamma \notag \\
&+& \Pi^{\sigma\beta} \Pi^{\delta \lambda} \left( \nabla_\beta R_{\delta \alpha}{}^\rho{}_{\gamma}- \tfrac 1 2 T^\epsilon_{\beta \delta} R_{\alpha \epsilon}{}^\rho{}_\gamma  \right) \chi_\rho \wedge \eta_\sigma \wedge \eta_\lambda \theta^\gamma\ ,\label{nablaReta}
\end{eqnarray}
modulo $\mathcal R^{\chi_\alpha}$, $\mathcal R^{\phi^\alpha}$ and $\mathcal R^{\theta^\alpha}$. The second term is easily seen to be zero from condition (\ref{conds2}).
To show the vanishing of the first set of terms,
we use the third condition in (\ref{conds}) and $\nabla_\alpha \Pi^{\lambda \beta}=-2 T^{[\lambda}_{\alpha \epsilon} \Pi^{ \beta] \epsilon}$  to compute
\begin{equation}
0=\nabla_\a (\Pi^{\lambda\beta} \nabla_\beta \widetilde R_{\gamma \delta}{}^{\rho \sigma})= \Pi^{\lambda \beta} (\nabla_\alpha \nabla_\beta \widetilde R_{\gamma \delta}{}^{\rho \sigma}+ T^\epsilon_{\alpha \beta} \nabla_\epsilon \widetilde R_{\gamma \delta}{}^{\rho \sigma}) \ .
\end{equation}
Employing the Ricci identity
\begin{equation*}
[\nabla_\alpha, \nabla_\beta] \widetilde R_{\gamma \delta}{}^{\rho \sigma}=-T^\epsilon_{\alpha \beta} \nabla_\epsilon \widetilde R_{\gamma \delta}{}^{\rho \sigma}+2 R_{\alpha\beta}{}^{(\rho}{}_\epsilon \widetilde R_{\gamma \delta}{}^{\sigma)\epsilon}+2R_{\alpha\beta}{}^\epsilon{}_{[\gamma} \widetilde R_{\delta] \epsilon}{}^{\rho\sigma} \ ,
\end{equation*}
one has
\bea
\Pi^{\lambda\beta}( \nabla_\beta \nabla_\alpha \widetilde R_{\gamma \delta}{}^{\rho \sigma} &+&2 R_{\alpha \beta}{}^{(\rho}{}_\epsilon  \widetilde R_{\gamma \delta}{}^{\sigma)\epsilon }+ 2 R_{\alpha \beta}{}^\epsilon{}_{[\gamma} \widetilde R_{\delta]\epsilon }{}^{\rho \sigma} )\notag \\
&=&\Pi^{\lambda \beta} (\nabla_\alpha \nabla_\beta \widetilde R_{\gamma \delta}{}^{\rho \sigma}+ T^\epsilon_{\alpha \beta} \nabla_\epsilon \widetilde R_{\gamma \delta}{}^{\rho \sigma} )=0 \ .
\eea
To show the vanishing of the third set of terms in (\ref{nablaReta}), we rewrite
the Bianchi identity $\nabla_{[\beta}R_{\delta \alpha]}{}^\rho{}_\sigma- T^\epsilon_{[\beta \delta} R_{\alpha]\epsilon}{}^\rho{}_\gamma=0$ as
\begin{equation}
2\left( \nabla_{[\beta} R_{\delta] \alpha}{}^\rho{}_\gamma -\tfrac1 2 T^\epsilon_{\beta\delta} R_{\alpha \epsilon}{}^\rho{}_\gamma  \right)+  \nabla_\alpha R_{\beta\delta}{}^\rho{}_\gamma - 2 T^\epsilon_{\alpha[\beta}R_{\delta]\epsilon}{}^\rho{}_\gamma=0 \ .
\end{equation}
On the other hand, the second condition in (\ref{conds}) together
with $\nabla_\alpha \Pi^{\sigma \beta}=-2 T^{[\sigma}_{\alpha \epsilon} \Pi^{ \beta] \epsilon}$ implies
\begin{equation}
0=\nabla_\alpha ( \Pi^{\sigma \beta} \Pi^{\delta \lambda} R_{\beta\delta}{}^\rho{}_\gamma )=\Pi^{\sigma \beta} \Pi^{\delta \lambda} (  \nabla_\alpha R_{\beta\delta}{}^\rho{}_\gamma - 2 T^\epsilon_{\alpha [\beta} R_{\delta]\epsilon}{}^\rho{}_\gamma ) \ .\label{id}
\end{equation}
Thus, contracting the above form of the Bianchi identity
by $\Pi^{\sigma \beta} \Pi^{\delta \lambda}$ and using (\ref{id})
it follows that the third term in $\mathcal A^{\eta_\alpha}$
vanishes as well.

In summary, we have showed that the universal Cartan integrability of the
equations of motion Eqs. (\ref{Rphi})--(\ref{Reta})
is equivalent to that the background fields $\Pi$
and $\Gamma$ can be used to define a differential
Poisson algebra.

\subsection{Gauge transformations}

\noindent Relying upon the framework for generalized
Poisson sigma models \cite{Park, Ikeda2001} (see also \cite{Ikeda2012, Boulanger2012, wip}), the universal
Cartan integrability of the equations of motion implies
that the action is invariant under Cartan gauge transformations.
On-shell, these transformations can be obtained
by first rewriting the equations of motion Eqs.
(\ref{Rphi})--(\ref{Reta}) on the canonical form
\begin{equation}
\widehat{\mathcal R}^i := dZ^i+ \widehat {\mathcal Q}^i(Z^j) = 0\ ,\qquad
Z^i:=(\phi^\a,\eta_\a;\theta^\a,\chi_\a)\ .
\end{equation}
Eliminating $d\phi^\a$ in $\nabla$ using ${\cal R}^{\phi^\a}=0$, we thus have
\begin{eqnarray}
\widehat {\cal R}^{\phi^\a}&=&d\phi^\a+ \Pi^{\a\b}\eta_\b \ ,\\[5pt]
\widehat {\cal R}^{\eta_\a}&=&d \eta_\a + \Pi^{\beta \gamma} \Gamma^\delta_{\beta \alpha}  \,\eta_\gamma \wedge \eta_\delta  +\Pi^{\gamma\lambda} R_{\a\gamma}{}^\b{}_\d \,\eta_\lambda
\wedge\chi_\b \theta^\delta+\tfrac{1}4  \nabla_\a \widetilde R_{\d\e}{}^{\b\gamma} \chi_\b \wedge \chi_\gamma \theta^\d\theta^\e \ , \qquad\\[5pt]
\widehat {\cal R}^{\theta^\a}&=&d \theta^\a-\Pi^{\beta \gamma} \Gamma^\alpha_{\beta \delta}  \eta_\gamma \theta^\delta + \tfrac 1 2 \widetilde R_{\gamma\d}{}^{\a\b} \chi_\b \theta^\gamma \theta^\d \approx 0\ ,\\[5pt]
\widehat {\cal R}^{\chi_\a}&=&d \chi_\a +
\Pi^{\beta \gamma} \Gamma^\delta_{\beta \alpha}  \eta_\gamma \wedge \chi_\delta - \tfrac 1 2  \widetilde R_{\a\d}{}^{\b\gamma} \chi_\b  \wedge \chi_\gamma \theta^\d \approx 0\ .
\end{eqnarray}
The on-shell gauge transformations are then given by
\begin{equation}
\delta Z^i=d \epsilon^i - \epsilon^j \frac{\partial}{\partial Z^j} \widehat {\mathcal Q}^i\ ,  \qquad
\mbox{modulo $\widehat {\cal R}^i$}\ ,
\end{equation}
where $\epsilon^i$ denote the gauge parameters, of which there is one
for each fields with strictly positive form degree, that is,
\be \epsilon^i=( 0, \epsilon^{(\eta)}_ \alpha;0, \epsilon^{(\chi)}_{\alpha} )\ ,\qquad {\rm deg}_2(\epsilon^i)=(-,0;-,0)\ ,\qquad \e_{\rm f}(\epsilon^i)=(-,0;-,1)\ .\ee
Thus, the infinitesimal gauge transformations are given by
\begin{eqnarray}
\delta \phi^\alpha&=&-\Pi^{\alpha \beta} \epsilon^{(\eta)}_{\beta}\ ,\\[5pt]
\delta \eta_\alpha &=& \nabla \epsilon^{(\eta)}_{\alpha}- \Pi^{\beta \gamma} \Gamma^{\delta}_{\beta \alpha}  \epsilon^{(\eta)}_\gamma \eta_\delta  -\Pi^{\gamma \lambda} R_{\a\gamma}{}^\b{}_\d \,\epsilon^{(\eta)}_\lambda \chi_\b \theta^\delta
+\Pi^{\gamma \lambda} R_{\a\gamma}{}^\b{}_\d \,\eta_\lambda
\epsilon^{(\chi)}_\b \theta^\delta\nn\\[5pt]&&-\tfrac 1 2  \nabla_\a \widetilde R_{\d\e}{}^{\b\gamma} \epsilon^{(\chi)}_\b \chi_\gamma \theta^\d\theta^\e\ ,\\[5pt]
\delta \theta^\alpha&=&   \Pi^{\beta \gamma} \Gamma^\alpha_{\beta \delta} \epsilon^{(\eta)}_{ \gamma} \theta^\delta  -\tfrac 1 2  \widetilde R_{\gamma \delta}{}^{\alpha \beta} \epsilon^{(\chi)}_{ \beta} \theta^\gamma \theta^\delta\ ,\\[5pt]
\delta \chi_\alpha  &=& \nabla \epsilon^{(\chi)}_\alpha-  \Pi^{\beta \gamma} \Gamma^\delta_{\beta \alpha} \epsilon^{(\eta) }_{\gamma}\chi_\delta+ \widetilde R_{\alpha \delta}{}^{\beta \gamma} \epsilon^{(\chi) }_{\beta} \chi_\gamma \theta^\delta \ ,
\end{eqnarray}
modulo ${\cal R}^i$.
Off-shell, it follows from the general formalism
\cite{Boulanger2012, wip}, that the gauge transformations
are given by
\be \delta Z^i=d \epsilon^i - \epsilon^j \frac{\partial}{\partial Z^j} \widehat{{\mathcal Q}}^i+
\tfrac12 \epsilon^k \,\widehat{\cal R}^l \,\partial_l \,\widehat\Omega_{kj} \,\widehat{\cal P}^{ji}\ ,\label{nic}\ee
where we have introduced the symplectic two-form
\be \widehat{\Omega}=d \widehat\Theta =\tfrac12 dZ^i \widehat{\cal O}_{ij} dZ^j=\tfrac12 dZ^i dZ^j \widehat\Omega_{ij}\ ,\qquad  \widehat{\cal P}^{ik}\widehat{\cal O}_{kj}=-\delta^i_j\ ,\ee
of degree three on the target space $\widehat N$ given in (\ref{widehatM}),
and the pre-symplectic structure
\be \widehat\Theta =\eta_\a \wedge d\phi^\a+\chi_\a \wedge \nabla \theta^\a\ ,\ee
treated as a one-form of $\mathbb N$-degree two on $\widehat N$.
Thus, the matrix $\widehat {\cal O}_{ij}$ can be read off from
\begin{equation}
\widehat \Omega= \tfrac1 2\begin{pmatrix}
d\phi^\rho & d\eta_\rho & d\theta^\rho & d\chi_\rho
\end{pmatrix}
\begin{pmatrix}
2 \partial_{[\rho} \Gamma^\alpha_{\gamma]\beta} \chi_\alpha \theta^\beta \quad &  \delta_\rho{}^\gamma \quad& -\Gamma^\alpha_{\rho \gamma} \chi_\alpha \quad& -\Gamma^\gamma_{\rho \alpha} \theta^\alpha \\
\delta^\rho{}_\gamma &  0 & 0 & 0 \\
-\Gamma^\alpha_{\gamma \rho} \chi_\alpha &  0 & 0 & -\delta_\rho{}^\gamma \\
\Gamma^\rho_{\gamma\alpha} \theta^\alpha &  0 & \delta^\rho{}_\gamma & 0
\end{pmatrix}
\begin{pmatrix}
d\phi^\gamma \\ d\eta_\gamma \\ d\theta^\gamma \\ d\chi_\gamma
\end{pmatrix}\ .
\end{equation}
Moreover, the components $\widehat{\cal P}^{ji}$ of the Poisson structure on $\widehat N$ is given by
\begin{equation}
\widehat{\cal P}^{ik}=\begin{pmatrix}
0 & - \delta^\sigma{}_\rho & 0 & 0 \\
-\delta_\sigma{}^\rho \quad &  R_{\sigma\rho}{}^\alpha{}_\beta \chi_\alpha \theta^\beta \quad& \Gamma^\rho_{\sigma\alpha}\theta^\alpha \quad & -\Gamma^\alpha_{\sigma\rho} \chi_\alpha \\
0 &  \Gamma^\sigma_{\rho\alpha} \theta^\alpha & 0 & -\delta^\sigma{}_\rho \\
0 &  \Gamma^\alpha_{\rho\sigma} \chi_\alpha & \delta_\sigma{}^\rho & 0
\end{pmatrix} \ .
\end{equation}
Using the above four by four matrices is simple to show that $\widehat{\cal P}^{ik}\widehat{\cal O}_{kj}=-\delta^i_j$.
If the connection vanishes identically, then the off-shell modification
of the gauge transformation \eqref{nic} vanishes.

\section{Conclusion and remarks}\label{Sec:4}

We have given an action of the covariant Hamiltonian form
that describes a two-dimensional topological sigma model
in a target space carrying the structure of a differential
Poisson algebra.
The kinetic term is given by the pull-back of a pre-symplectic
form that is non-canonical and hence the off-shell gauge
transformations contain an additional set of terms 
proportional to the Cartan curvatures. 
Besides the characteristic local symmetries of such models,
whose requirements are indeed equivalent to those of the
Jacobi identities of the differential Poisson algebra,
our action also exhibits a rigid supersymmetry corresponding
the de Rham differential on the Poisson manifold.
This rigid symmetry fixes the coefficient of the quartic fermion
coupling.
We expect that the AKSZ quantization
\cite{Cattaneo, Cattaneo:2001ys} of the original Poisson sigma model
can be generalized to the present model in a background
diffeomorphism covariant fashion, \emph{i.e.} such that
there exists a generalization of \eqref{diff} to the
gauge fixed action.
Assuming furthermore that Kontsevich's formality theorem
generalizes to the deformation of the graded
Poisson bracket,
the similarity between the structures of the Poisson bracket \eqref{PB} and the action \eqref{action}
suggests that the correlation functions of suitable
boundary vertex operators yield the covariantized version
of Kontsevich star product (at least in simple target
space topology).

Clearly, the first steps in this direction are to reproduce
the generalized Poisson bracket (\ref{PB}) at order
$\hbar$ and then verify the bi-differential operator
found in \cite{Tagliaferro,Zumino} at order $\hbar^2$,
which we leave for separate considerations\footnote{
Starting from a path integral
weighted by $\exp(\frac{i}\hbar S)$, the perturbative
expansion is obtained by rescaling $(\eta_\a,\chi_\a,\theta^\a)\rightarrow (\hbar \eta_\a,\sqrt{\hbar}\chi_\a,\sqrt{\hbar}\theta^\a)$
and expand around the two-dimensional vacuum in
which $\langle \phi^\a\rangle$ and $\langle \theta^\a \rangle$ are
constant and $\langle \eta_\a\rangle$ and $\langle \chi_\a \rangle $
vanish.}.
More precisely, we propose that the BRST cohomology
of the model contains a ring generated by the zero-modes of
$(\phi^\a,\theta^\a)$ that realizes the star product
deformation of the space $\Omega(N)$ of differential
forms on $N$.
As already mentioned in the Introduction and using the
notation of Section \ref{Sec:3.1},
one can map the elements $d\phi^{\a_1}\wedge\cdots \wedge d\phi^{\a_p}\omega_{\a_1\dots \a_p}$ in $\Omega(N)$ to
elements $\theta^{\a_1}\cdots \theta^{\a_p}
\omega_{\a_1\dots \a_p}$ in the subspace $\Omega_{[0]}(T[0,1]N)$
of zero-forms in the space $\Omega(T[0,1]N)$ of
differential forms on $T[0,1]N$.
Likewise, in the gauged-fixed theory\footnote{In the gauge
fixed theory all quantities are assigned form degrees,
ghost numbers and additional Grassmann parities,
and we choose the Koszul sign convention to be given by
$AB=(-1)^{
|A||B|+\e_{\rm f}(A)\e_{\rm f}(B)} BA$ where
the total degree $|A|={\rm deg}_2(A)+ {\rm gh}(A)$.
}, the ghosts $(c_\a,\gamma_\a)$ for $(\eta_\a,\chi_\a)$
have form degree zero, ghost number one and
additional Grassmann parities $\e_{\rm f}(c_\a,\gamma_\a)=(0,1)$.
Their zero-modes yield realizations of star product deformations
of the spaces $Poly^{(\pm)}(N)$ of symmetric $(+)$
and anti-symmetric $(-)$ polyvector fields on $N$ by mapping
anti-symmetric poly-vectors
$\Pi^{\a_1\dots \a_n}(\phi) \partial_{\a_1}\wedge \cdots \wedge \partial_{\a_n}$ to $\Pi^{\a_1\dots \a_n}(\phi) c_{\a_1}\cdots c_{\a_n}$ in
$\Omega_{[0]}(T^\ast[1,0] N)$
and symmetric polyvectors $G^{\a_1\dots \a_n}(\phi) \partial_{\a_1}\odot \cdots \odot \partial_{\a_n}$
to $G^{\a_1\dots \a_n}(\phi) \gamma_{\a_1}\cdots \gamma_{\a_n}$
in $\Omega_{[0]}(T^\ast[1,1] N)$.
It would be interesting to examine the resulting
target space quantum geometries in more detail.

The action \eqref{action}, which describes a classically
topological theory that remains to be gauge fixed,
bears a close resemblance to the complete action of
the first order formulation \cite{Baulieu,Frenkel,Bonechi}
of the topological A model \cite{Witten}.
The latter is obtained by a
topological twist of the $\mathcal N=(2,2)$
supersymmetric sigma model, and
requires the target space to be K\"ahler,
and hence symplectic, unlike our model,
whose target space is only required to
be a Poisson manifold.
Moreover, the type A model refers to
a worldsheet metric, which enters via additional
couplings to the hermitian metric and its
compatible curvature of the form
$g^{\a\b} \eta_\a\wedge\ast\eta_\b$
and $g^{\a\e}R_{\gamma \d}{}^{\b}{}_\e(g) \chi_\a\wedge\ast \chi_\b
\theta^\gamma \theta^\d$.
Thus the type A model action is non-singular in the
sense that it does not admit any local symmetries.
Instead, the couplings are tuned such that
the complete action is exact under a rigid nilpotent
supersymmetry\footnote{
The rigid supersymmetry generator
of the A model is sometimes
referred to as a BRST operator,
even though the twisting is not
a gauge fixing procedure.
Attempts to identify the
type A model as a gauge fixed
version of a classically topological
theory have been made in
\cite{BaulieuSinger}.},
whose factorization yields a topological model.
Our model, on the other hand, is classically
topological without requiring the classical observables
to be $\delta_{\rm f}$-closed.
Thus, in the terminology of topological field theories,
our model is of the Schwarz type, while the A model
is of the Witten, or cohomological, type.

It would be interesting to examine
whether there are more robust relations between
the type A and B models and also the interpolating
A-I-B model \cite{Frenkel}, including their infinite
(and possibly zero) volume limits, and our model and
various deformations of it.
As for the latter, one may consider adding Yukawa couplings
formed out of the $S$-tensor defined in \eqref{S}
and additional metric couplings $G^{\a\b}\chi_\a\wedge \chi_\beta$
(which add terms of intrinsic degree minus two to the bracket).
One may also seek ways to couple of our model to two-dimensional gravity,
which may be of importance for the
formulation of the theory on worldsheets
of higher genus, and possibly new topological open strings.
To this end, besides exploring the relations to
the type A and B models, it may also be
fruitful to explore another route, based on
the observation that prior to
adding the worldsheet fermions, the Poisson
sigma model exhibits vacuum bubble cancelations
in simple worldsheet topologies.
Adding the fermions lead to that these cancellations
generalize to arbitrary topologies \cite{Wu} (including
boundaries).
Thus, including bubbles with external  matter legs,
one may expect anomaly-induced topological matter-gravity couplings.
We plan to address these issues in a future work.

One motivation behind the present work
is Vasiliev's higher spin gravity, whose field theoretic
formulation is in terms of differential
star product algebras \cite{Vasiliev}.
The explicit models that have been constructed
so far are formulated on products of commuting
manifolds, containing spacetimes, and symplectic
manifolds of simple topology, quantized using the
Moyal star product.
The covariantized Kontsevich formalism provides
a tool facilitating the formulation of higher
spin gravities on manifolds of more
general topology, possibly as Frobenius--Chern--Simons
theories (or BF analogs thereof) following \cite{fcs}.
Its extension to topological open strings, with
non-trivial topological expansions, may lead
to complementary first-quantized descriptions
of higher spin gravity.
The latter perspective is supported by the recent
progress in computing higher spin tree amplitudes
starting from traces over oscillator algebras
\cite{Colombo1,Colombo2,Didenko1,Didenko2},
in its turn motivated by the proposal
made in \cite{Engquist:2007pr} for how Vasiliev's
theory arise in tensionless limits of
closed strings in anti-de Sitter spacetime.

Finally, a natural part of the application
to higher spin gravity as well as discretized
strings, and also more general
constrained systems, is the gauging of
Killing symmetries of our Poisson sigma model.
In principle, this procedure ought to be straightforward
and leads to a natural generalization of the
original gauged Poisson sigma model,
which we expect to report on in a forthcoming
publication.

\paragraph{Acknowledgement.}
We are thankful to Francisco Rojas, Ergin Sezgin and Brenno Vallilo
for discussions.
A.T.G. is supported by FONDECYT post-doctoral grant number 3130333. P.S. is supported by Proyecto Conicyt DPI20140115. N.B. is a Research Associate of the Fonds de la Recherche Scientifique$\,$-FNRS (Belgium).
His work was partially supported by the contract ``Actions de Recherche concert\'ees$\,$-Communaut\'e francaise
de Belgique'' AUWB-2010-10/15-UMONS-1. C.A. is supported by a UNAB PhD scholarship.

\appendix
\section{Conventions and notation}\label{conventions}

\noindent The covariant exterior derivatives of the components of a vector field $V=V^\alpha \partial_\alpha$ and a one-form $\omega=\omega_\alpha d\phi^\alpha$ are given by
\be
\nabla V^{\a} = dV^{\a}+ \C^\a{}_\b V^{\b} \ ,\qquad
\nabla \omega_{\a} = d\omega_{\a} - \C^\b{}_\a \wedge \omega_{\b}\ ,\ee
where $\Gamma^\a{}_\b=d\phi^\gamma \Gamma^{\a}_{\gamma \b}$ is the connection one-form. In terms of components, we have $\nabla V^\a= d\phi^\beta \nabla_\b V^\a$ and $\nabla \omega_\a=d\phi^\beta \nabla_\b \omega_\a$
where
\be
\nabla_{\a} V^{\b} = \partial_{\a} V^{\b} + \C_{\a \gamma}^{\b} V^{\gamma}\ ,\qquad
\nabla_{\a} \o_{\b} = \partial_{\a} \o_{\b} - \C_{\a \b}^{\gamma} \o_{\gamma}\ .
\ee
The basic Ricci identities read
\be
[\nabla_{\a}, \nabla_{\b}] V^{\gamma} = - T_{\a\b}^{\d} \nabla_{\d} V^{\gamma}+R_{\a\b}{}^{\gamma}{}_{\d} V^{\d} \ ,\qquad [\nabla_{\a}, \nabla_{\b}] \o_{\gamma} = - T_{\a\b}^{\d} \nabla_{\d} \o_{\gamma}-R_{\a\b}{}^{\d}{}_{\gamma} \o_{\d} \ ,
\ee
where the curvature and torsion tensors
\be
 R_{\a\b}{}^{\gamma}{}_{\d}= 2\, \partial_{[\a} \C_{\b]\d}^{\gamma} + 2\, \C_{[\a|\e}^{\gamma} \C_{|\b]\d}^{\e}  \ ,\qquad
 T_{\a\b}^{\gamma} = 2\, \C_{[\a\b]}^{\gamma}\ .
\ee
The corresponding curvature and torsion two-forms
\be R^\a{}_\b=\tfrac12 d\phi^\gamma \wedge d\phi^\d R_{\gamma\d}{}^\a{}_\b\ ,\qquad T^\a=\tfrac12 d\phi^\gamma \wedge d\phi^\d T_{\gamma\d}^\a\ ,\ee
which can also be written as
\be R^\a{}_\b = d\C^\a{}_\b + \C^\a{}_\gamma \wedge \C^\gamma{}_\b\ ,\qquad T^\a=\C^\a{}_\b \wedge d\phi^\b\ .\ee
The covariant exterior derivative of the one-form itself is given by
\be d\omega=\nabla\omega = (\nabla d\phi^\a) \omega_\a+ (\nabla \omega_\a) d\phi^\a = d\phi^\a d\phi^\b (\nabla_\a \omega_\b+\tfrac12 T^\gamma_{\a\b} \o_\gamma)
\ .\ee
The Bianchi identities read
\be T^\a=\nabla d\phi^\a\ ,\quad \nabla T^\a= R^\a{}_\b \wedge d\phi^\b\ ,\qquad \nabla R^\a{}_\b = 0 \ ,\ee
or in components
\be
R_{[\a\b}{}^{\gamma}{}_{\d]} = \nabla_{[\a} T_{\b\d]}^{\gamma} - T_{[\a\b}^{\e} T_{\d] \e}^{\gamma}\ ,\qquad \nabla_{[\a} R_{\b\gamma]}{}^{\d}{}_\e- T^\lambda_{[\a\b} R_{\gamma] \lambda}{}^\d{}_\e=0\ .\ee
The square of the exterior covariant derivative acting on the components of
a vector field and a one-form are given by
\be \nabla^2 V^\a = R^\a{}_\b V^\b\ ,\qquad \nabla^2 \omega_\a= - R^\b{}_\a \wedge \omega_\b\ .\ee
In analyzing the differential Poisson algebra, it is convenient
to define a new connection $\widetilde \nabla$ with connection
coefficients $\widetilde \Gamma^\alpha_{\gamma \beta}:=\Gamma^\alpha_{\beta \gamma}$.
We make repeated us of the identity
\be \nabla_\a \Pi^{\b\gamma}= \widetilde \nabla_\alpha \Pi^{\beta \gamma}-2T^{[\b}_{\a\d} \Pi^{\gamma]\d}\ .\ee
We denote the components of the curvature of $\widetilde \nabla$
by $\widetilde R_{\alpha \beta}{}^\gamma{}_\delta$ and define
\be \widetilde R^{\a\b}:= \Pi^{\b\gamma} \widetilde R^{\a}{}_\gamma  \ .\ee


\end{document}